# Observational evidence that a Gaia-type feedback control system with proportional-integral-derivative characteristics is operating on atmospheric surface temperature at global scale

L. Mark W. Leggett* & David A. Ball

Global Risk Policy Group Pty Ltd, Townsville, QLD, Australia

*mleggett.globalriskprogress@gmail.com

September 26 2018

**The Gaia hypothesis (Lovelock and Margulis, 1974) proposes that there is a control system operating at global level that regulates climate and chemistry at a habitable state for the biota. Here we provide statistically significant observational evidence that a feedback control system moderating atmospheric temperature is presently operating coherently at global scale – that is to say, observational evidence for Gaia. Further, this control system is of a sophisticated type, involving the corrective feedback not only of a linear error term but also its derivative and its integral. This makes it of the same type as the most commonly used control system developed by humans, the proportional-integral-derivative (PID) control system.**

In the fifty years since its inception, the Gaia hypothesis has been given a number of meanings. Its core meaning, however, has remained unchanged. Lovelock and Margulis (1974) wrote: "The notion of the biosphere as an active adaptive control system able to maintain the Earth in homeostasis we are calling the 'Gaia' hypothesis..."A more recent version cited favourably by Wilkinson (2014) is from Lovelock (2003): "…organisms and their material environment evolve as a single coupled system, from which emerges the sustained self-regulation of climate and chemistry at a habitable state for whatever is the current biota'."

With the term "control system able to maintain the Earth in homeostasis" being used in 1974 and "self-regulation for the biota" being used in 2014, we see Gaia being consistently described as *a control system operating at global level and regulating climate among other factors at a habitable state for the biota*, and that is the definition we will use here.

We now turn to the current state of the question of evidence of Gaia. In 2012, Moody would still write: "Does there in fact exist an emergent property of life that operates in a manner conducive to the continuation of the system as a whole? This is a question suitable for empirical research…"

Dyke and Weaver (2013) wrote: "Despite such criticism, research into the plausibility of Gaian homeostasis has continued in the context of: natural selection; ecology and evolution; biogeochemical regulation; and complex adaptive systems. Notwithstanding this work the hypothesis is still far from being confirmed."

These references suggest that empirical evidence for Gaia - as a global-level control system – is still outstanding.

We open this discussion by firstly reviewing control systems, then explaining the concept of Gaia, and finally looking at how to test the hypotheses relating to Gaia.

Control systems are in widespread everyday use, from the cruise control in a motor vehicle to the thermostat used for building heating/air-conditioning systems, among many other applications.

In a control system (Astrom and Murray, 2008) a *controller* is used to automatically adjust a *controller output* so as to hold the trend over time of the variable of interest (termed the *process variable*) at its *setpoint*. The setpoint is the value of the preferred level for the process variable time series. A factor driving the process variable away from its setpoint is termed a *disturbance*. The *error* is defined as the difference between the level of the setpoint and the level of the disturbance.

To describe the domestic room thermostat in these terms (VanDoren, 1998), the thermostat is the controller, the actual room temperature is the process variable, the desired room temperature is the setpoint, the activation signal to the air conditioner or heater is the controller output, and random heat sources (such as radiant heat from sunshine and warm bodies, or radiant cooling through windows) constitute the disturbances to the process.

The most common general-purpose controller uses a control-loop feedback mechanism. In this, it has been found that feeding back the *derivative* and the *integral* of the error term as well as the *level* of the error term gives more accurate control than simply feeding back the error term. These two extra terms reduce the problems of overshoot and undershoot of the variable being controlled. Based on the nature of the three feedback terms used, this control system is called a proportional-integral-derivative (PID) controller (Araki, 2002; Astrom and Murray, 2008).

We now turn to the Gaia-type control system and to the fact that both its conceptualisation and testability have been questioned. The two main authors who have attempted structured and systematic assessments of the Gaia hypothesis are Kirchner (1989, 1991, 2002, 2003), and Kleidon (2002). Tyrrell (2013) has also published an assessment in book form.

Kirchner's and Tyrrell's analyses are critiques. Kirchner makes points under two headings: ways of conceptualising Gaia; and ways of setting up hypotheses to test for the existence of Gaia. We develop our method using these two headings.

First, on conceptualising Gaia, Kirchner (2002) refers to three major themes of the Gaia hypothesis and provides critiques of them: Gaia and Homeostasis, Gaia and Environmental Enhancement, and Gaia and Natural Selection. The first two themes are germane to this paper.

With regard to Gaia and Homeostasis Kirchner (2002) asked whether biological feedbacks stabilise the global environment. That is, is the 'Homeostatic Gaia' hypothesis correct?

Kirchner (2003) argues that:

> in the context of the past 30 years …Gaia's proponents have vigorously advanced the view that life stabilizes Earth's climate, while largely ignoring the substantial body of evidence that biological feedbacks also can (and do) destabilize Earth's climate. It is also a fair question in the context of Lenton (2002), despite Lenton and Wilkinson (2003)'s assertion that 'an effort was made to objectively balance available evidence'. While Lenton clearly acknowledges the theoretical possibility of biologically mediated positive feedbacks, in 14 pages he makes only a few passing references to the biological feedbacks that actually do destabilize Earth's climate.

In response to this point as earlier presented in Kirchner (2002), Lenton and Wilkinson (2003) wrote:

> It has been proposed that on average the biota alters the physical and chemical environment in a manner that benefits them (without this being teleological). Gross primary productivity … and 'cycling ratios' … have been suggested as metrics of this. …
>
> We suggest that the dominant process should be apparent when the life environment system is perturbed or collapses. A system with strong environmental feedback will be prone to rapid transitions between states, whereas one where adaptation dominates will change more gradually. This has been observed in two-dimensional variants of the Daisyworld model...
>
> The rainforest … may be a good real world test case.

Kirchner (2003) later went on to say:

> I have been arguing for more than a decade that the quest for grand generalizations is *part of the problem*, because it can blind us to the diversity of feedback processes in the Earth system (Kirchner, 1989). I certainly agree with LW that, 'We should not expect there to be universal truths about the behavior of such a complex system . . . Life is not always going to enhance gross primary productivity . . . or any other metric of the system'. But I disagree with their view that 'What matters is the balance of examples: do they suggest a tendency in one direction?' In my view, what matters is not 'the balance of examples', what matters is *how the Earth system works*…

The present authors agree with Kirchner that generalising from a range of sub-global examples is not a compelling answer to the question of whether life stabilises Earth's climate at full global scale. However we would contest whether finding out "how the Earth system works" is a sufficiently precise alternative approach.

We consider that the appropriate approach is to look at a *global level* performance metric and see if its trend is towards stabilisation or destabilisation.

Lenton and Wilkinson (2003) refer to such global level performance and global level metrics when they write: "It has been proposed that on average the biota alters the physical and chemical environment in a manner that benefits them … Gross primary productivity … and 'cycling ratios' …  have been suggested as metrics of this."

However when suggesting when "the dominant process should be apparent", they first refer to results achieved from a simulation rather than from observations ("This has been observed in two-dimensional variants of the Daisyworld model"). It should be noted that simulations can clarify and help the formulation of ideas. However only observations can provide evidence for them (Karplus, 1977, 1992;, Enting, 1987; Leggett and Ball, 2018). When Lenton and Wilkinson (2003) move to discussing observations, it is only to potential results and then only from a subset of the planet ("The rainforest … may be a good real world test case").

Within the Gaia hypothesis, one of the essential processes for life that is considered to be regulated by such a control system is the global atmospheric surface temperature (Lovelock and Margulis, 1974). We therefore use global surface temperature as the global level performance metric, specifically the HadCRUT4 data set (Morice et al., 2012).

We now turn to Kirchner's Gaia and Environmental Enhancement theme, which he describes as "the claim that the biota alter the physical and chemical environment to their own benefit."

This appears to ask if a move to any new stabilisation leads to life in aggregate being better off.  In this study, we address this question by assessing whether the trend in the global level performance metric that we use, global surface temperature, is towards stabilisation or destabilisation, and then asking whether the new state is more or less beneficial to the biota.

We turn now to the series of points which Kirchner (1989) outlined on hypotheses and their testing.

In his paper, Kirchner (1989) wrote:

> The day-to-day business of science consists of testing hypotheses, but some hypotheses cannot be tested. …
>
> The minimal criteria of testability can be stated concisely. In order to be testable, a hypothesis must be clear, and its terms must be unambiguous.
>
> It must be intelligible in terms of observable phenomena.
>
> And most importantly it must generate predictions of two kinds: confirmatory predictions (phenomena that should be observed if the hypothesis is true and that would not be predicted by the existing body of accepted theory) and falsifying predictions (phenomena that should be observed if the hypothesis is false).

Our results are based on observations, and as will be seen, demonstrate the characteristics of a Proportional-Integral-Derivative control system. The patterns in the observations that we present are not predicted by any rival theory.

We assert that predictions arising from rival theories are not relevant to whether or not a prediction of the theory being assessed is confirmatory. However whether the theory in question makes predictions not made by any rival theory is important for evaluating a new theory. The control-system-related patterns in the observations that we present are not predicted by any rival theory.

As far as falsifying predictions are concerned, if there is no control system then global surface temperature will not follow control-system-related patterns.

Further on hypothesis testing, use of the randomised controlled trial is often considered the gold standard for the provision of scientific evidence of cause and effect (for example, see Sackett et al., 1996). The randomised controlled trial requires a subject group and a control group, which for Gaia would mean a planet in one instance with Gaia and in another without. For example, Lenton (2002) wrote:

> Testing hypotheses for the effect of life on the Earth system (including potential gross primary productivity (Kleidon, 2002)) is difficult because we are unable to observe the Earth in a 'dead' state, except by sterilising the planet and waiting millions of years for it to approach an abiotic steady state (which is undesirable, to say the least!).

Use of a subject entity and a control entity, however, is not the only way to test hypotheses. Another form of scientific theory testing involves (Bogen, 2017): "comparing observation sentences describing observations made in natural or laboratory settings to observation sentences that should be true according to the theory to be tested."

This method is one supported by Kirchner (1989) as outlined above ("And most importantly it must generate …confirmatory predictions (phenomena that should be observed if the hypothesis is true…)").

One reaction to the testability point has been the development and use of the Daisyworld simulation (Lenton and Wilkinson, 2003). But as argued above, simulations can indicate where to look for evidence but can not provide evidence. Observations from nature are needed to provide evidence.

Observational evidence showing the existence of a control process operating now, and at a global scale, would provide strong evidence for Gaia.

This is the hypothesis for which we test. As mentioned, we have attempted to ensure our hypothesis testing method is in line with that of Kirchner (1989) and Bogen (2017).

In Leggett and Ball (2015), we reported that the three major climate series – for atmospheric pressure (the example we used was the Southern Oscillation Index (SOI)); the global surface temperature; and the level of atmospheric $CO_2$ – related to each other such that atmospheric $CO_2$ was the integral of temperature and SOI was the derivative of temperature.

Here we recognise that this identity is precisely analogous to that found in the PID control system.

PID systems have been proposed theoretically to exist for living systems, including those influencing climate. For example, Nisbet (2012) wrote:

> Figure 3, modelled on industrial PID (proportional–integral–derivative) controllers, illustrates the way natural selection, acting on rubisco specificity, may have managed the surface temperature, at least in the past ~2.3 Ga since the air became oxygen-rich. The "optimum" temperature is the temperature at which the contemporaneous (at that "present") global biosphere has maximum productivity. If an external perturbation occurs, such as a volcanic eruption or solar warming, then there will be a proportional response as photosynthesis and respiration increase/decrease or decrease/increase, taking up or releasing $CO_2$. Methane too will change, especially as temperature change affects water precipitation in rain and snow.

This paper, then, seeks evidence for the current existence and operation of a control system of PID type affecting the global-level indicator, atmospheric surface temperature.

To this end we next seek to conceptualise the formal control system terms from the perspective of a putative control system for global surface temperature.

Once this is achieved, testing is then done for both the full PID type of control system and also for simpler versions using subsets of the P, I and D terms.

**Methods**

Statistical methods used are standard (Greene, 2012) and generally as used in Leggett and Ball (2015). Categories of methods used are normalisation; differentiation (approximated by differencing); integration (approximated by the cumulative sum); and time-series analysis. Within time-series analysis, methods used are: Z-scoring; smoothing; leading or lagging of data series relative to one another to achieve best fit; testing for the order of integration of each series (a prerequisite for using data series in time-series analysis); dynamic regression modeling to include autocorrelation in models; and ridge regression to address collinearity. These methods will now be described in turn.

**Z-scoring of series.** To make it easier to visually assess the relationship between the variables, the data were normalised using statistical Z scores or standardised deviation scores (expressed as "relative level" in the figures). In a Z-scored data series, each data point is part of an overall data series that sums to a zero mean and variance of 1, enabling comparison of data having different native units. Hence, when several Z-scored time series are depicted in a graph, all the time series will closely superimpose, enabling visual inspection to clearly discern the degree of similarity or dissimilarity between them. Individual figure legends contain details on the series lengths used as base periods for the Z-scoring.

A regression using Z-scored variables provides standardised regression coefficients. These coefficients report how much change a one-standard-deviation change in the independent variable produces in the dependent variable. Although comparisons between these coefficients must be interpreted with care, a standardised coefficient for independent variable *a*, for example, of 2 indicates that independent variable *a* is twice as influential upon the dependent variable as another independent variable that has a standardised coefficient of 1 (Allen 1997).

In the time-series analyses, global atmospheric surface temperature is the dependent variable. We tested the relationship between this variable and independent variables derived from (1) the level of atmospheric $CO_2$; (2) its change; and (3) its cumulative sum. We express the change in $CO_2$ as its first finite difference (we label this "first difference" in the text). Variability is explored using intra-annual (monthly) data. The period covered in the figures is shorter than that used in the data preparation because of the loss of some data points due to calculations of differences and of moving averages.

**Smoothing.** Smoothing was used on series incorporating first-difference $CO_2$ to the degree needed to reveal significant relationships if any were present. Smoothing is carried out initially by means of a 13-month moving average – this also minimises any remaining seasonal effects. In the following analyses, as for Leggett and Ball (2015), it is found that the d_ $CO_2$ (D_error) term fits the regression best when smoothed by a 13-month moving average.

**Leading or lagging.** Variables are led or lagged relative to one another to achieve best fit. These leads or lags were determined by means of time-lagged correlations (correlograms). The correlograms were calculated by shifting the series back and forth relative to each other, one month at a time.

**Labelling data series.** With this background, the convention used in this paper for unambiguously labelling data series and their treatment after smoothing or leading or lagging is depicted in the following example:

The atmospheric $CO_2$ series is transformed into its first difference and smoothed with a 13-month moving average. The resultant series is then Z-scored using the period 1958 to 1976 as a base. The smoothed series is found to fit temperature best with a one month lead. The resulting series so led is expressed as 13mma_d_$CO_2$_Z5876.

Note that to assist readability in text involving repeated references, atmospheric $CO_2$ is sometimes referred to simply as $CO_2$ and global surface temperature as temperature.

**Time series analysis.** Time series models (Greene 2012) differ from ordinary regression models in that the results are in a sequence. Hence, the dependent variable cannot be influenced only by the independent variables, but also by prior values of the dependent variable itself. This is termed autocorrelation between measured values.

This serial nature of the measurements must be addressed by careful examination of the lag structure of the model. This type of ordinary least squares regression is termed 'time series analysis' (Greene, 2012).

A further issue in time series analysis concerns what is termed the 'order of integration' of each of the series used. Greene (2012) states: "The series yt is said to be integrated of order one, denoted I(1), because taking a first difference produces a stationary process. A non-stationary series is integrated of order d, denoted I(d), if it becomes stationary after being first-differenced d times. An I(1) series in its raw (undifferenced) form will typically be constantly growing, or wandering about with no tendency to revert to a fixed mean."

The usual tests for stationarity (for example, the ADF test) can allow for the presence of a linear deterministic trend in the time-series in question, but they cannot allow for the possibility of a polynomial trend. It will be shown that this possibility needs to be tested for in this study.

There is evidence that if a unit root test that allows for only a linear trend is applied mistakenly to a series with a polynomial trend, then the test may have extremely low power. For example, see the results of Harvey et al. (2008) in the context of the de-trended ADF-type tests of Elliott et al. (1996).

This problem may be resolved by using the Lagrange multiplier (LM) unit root test(s) proposed by Schmidt and Phillips (1992). These tests make explicit allowance for the possibility of a polynomial deterministic trend of order up to 4 in the series under test. As with the ADF test, the null hypothesis is that the series has a unit root, and the alternative hypothesis is that it is stationary. A value for the LM test statistic that is more negative that the tabulated critical value leads to a *rejection* of the null hypothesis, and suggests that the series is stationary.

The Schmidt-Phillips test is available in the urca package in the R statistical software (R Development team, 2009) and is described in the documentation by Pfaff et al. (2016). Specifically, the ur.sp-class allows us to apply the Schmidt-Phillips tests. Although there are two such tests, the so-called $\tau$ test and the $\rho$ test, the results of only the former test are reported. (The results of the $\rho$ test led to exactly the same conclusions.) Testing is reported at the 1% significance level, but the results are not sensitive to this choice.

Next, a model must be established in which any autocorrelation in the short-run relationship, if present, is fully accounted for by use of an optimal lag structure. In this study, this is done within the modelling process by reference to the adjusted coefficient of determination $R^2$ and to the Akaike Information Criterion.

The adjusted $R^2$ is used because (Greene, 2012) there is a problem that the unadjusted $R^2$ cannot fall when, in multiple regression, independent variables are added to a model so there is a built-in tendency to add more independent variables to the model. To address this issue, the adjusted $R^2$ is a fit measure that penalises the loss of degrees of freedom that result from adding variables to the model. There is, however (Greene, 2012) some question about whether the penalty is sufficiently large to ensure that the criterion will necessarily lead the analyst to the correct model as the sample size increases. Several alternative fit measures termed 'information criteria' have been developed. Several criteria are available for selection within the Eviews ARDL method. We use the ARDL default criterion, the Akaike criterion.

Pesaran et al. (2001) point out that ARDL modelling is "also based on the assumption that the disturbances … are serially uncorrelated. It is therefore important that the lag order p of the underlying VAR is selected appropriately. There is a delicate balance between choosing p sufficiently large to mitigate the residual serial correlation problem and, at the same time, sufficiently small so that the conditional ECM is not unduly over-parameterized, particularly (when) limited time series data … are available."

To avoid over-parameterisation, we seek the model with the fewest lags that produces a model with no autocorrelation out to 36 lags.

The specific information sought from each well-specified ARDL model is the effect on the dependent variable of each of the potential driving variables, the relative percentage of the total driving task that each achieves, and the degree of statistical significance of each.

The ARDL econometric model used here was that implemented in the Eviews 9.5 package (IHIS Eviews, 2017). The time series statistical software package Gnu Regression, Econometrics and Time-series Library (GRETL) was also used.

**Collinearity.** Intrinsic to its design, a control system which involves some or all of proportional, integral and derivative feedback terms creates a problem for its analysis by means of ordinary least squares (OLS) econometric time-series methods. The problem is that one independent variable, the proportional (linear) error (P error) variable – albeit displaying by further design the opposite sign – will be exactly collinear with another variable – the Disturbance variable.

On collinearity (interdependency), Farrar and Glauber (1964) write:

> The single equation, least squares regression model is not well equipped to cope with interdependent explanatory variables. In its original and most simple form the problem is not even conceived. Values of X are presumed to be the pre-selected controlled elements of a classical, laboratory experiment. Least squares models are not limited, however, to simple, fixed variate – or fully controlled – experimental situations. (Other situations) may provide the data on which perfectly legitimate regression analyses are based.

Hastie et al. (2009, p51) write:

> The Gauss–Markov Theorem: One of the most famous results in statistics asserts that the least squares estimates of the parameters $\beta$ have the smallest variance among all linear unbiased estimates. We … make clear that the restriction to unbiased estimates is not necessarily a wise one.

On a problem arising from OLS based on the Gauss–Markov theorem, Greene (2012, p129) writes:

> …the Gauss–Markov theorem states that among all linear unbiased estimators, the least squares estimator has the smallest variance. Although this result is useful, it does not assure us that the least squares estimator has a small variance in any absolute sense. Consider, for example, a model that contains two explanatory variables and a constant. … If the two variables are perfectly correlated, then the variance is infinite.

The problem is that variance being infinite means that standard OLS packages will not run.

We now turn to solutions. Hastie et al. (2009, p51) write:

> The Gauss-Markov theorem implies that the least squares estimator has the smallest mean squared error of all linear estimators with no bias. However, there may well exist a biased estimator with smaller mean squared error. Such an estimator would trade a little bias for a larger reduction in variance.
>
> …most models are distortions of the truth, and hence are biased; picking the right model amounts to creating the right balance between bias and variance.
>
> Biased estimates are commonly used. Any method that shrinks or sets to zero some of the least squares coefficients may result in a biased estimate. (These include) variable subset selection and ridge regression…

On variable subset selection Greene (2012, p131) writes: "The obvious practical remedy (and surely the most frequently used) is to drop variables suspected of causing the problem from the regression…" We use this approach as one method in what follows.

Ridge regression takes another approach to reducing a variance that would otherwise be infinite, by introducing bias in a controlled way into the regression. This enables the variables in question to stop displaying infinite variance, and so the assessment of the performance of a model including the full set of independent variables is possible. Ridge regression (Hoerl, 1962; Birkes and Dodge, 1993) is one member of a family of methods which achieve this by imposing a penalty on ("shrinking") the size of relevant least squares coefficients. In this study, the Gretl ridge regression function (Schreiber, 2017) is used. The method requires an optimal shrinkage parameter (ridge constant or lambda) to be determined. This is done by use of a ridge trace process (Hoerl and Kennard,1970) in which estimated coefficients are compared against a range of shrinkage parameters in order to seek the most favorable trade-off of bias against precision (inverse variance) of the estimates.

**Data.** For global surface temperature, we used the Hadley Centre–Climate Research Unit combined Landsat and SST surface temperature series (HadCRUT) version 4.6.0.0 (Morice et al., 2012). In the tables, figures and text in the paper, this series is termed 'global surface temperature'. For atmospheric $CO_2$ data from 1958 to the present, the US Department of Commerce National Oceanic and Atmospheric Administration Earth System Research Laboratory Global Monitoring Division

Mauna Loa, Hawaii, annual $CO_2$ series (Keeling et al., 2009) is used. In the paper, this series is termed 'atmospheric $CO_2$'.

In the analyses, monthly data is used. The period covered in the figures is sometimes shorter than that used in the data preparation because of the loss of some data points due to calculations of differences and of moving averages.

We note that to assist readability in text involving repeated references, atmospheric $CO_2$ is sometimes referred to simply as '$CO_2$' and global surface temperature as 'temperature'. The time period covered by the time series used in each table or figure is given in the title of the table or figure.

In the tables of results, statistical significance at the 10%, 5% and 1% levels is indicated by the symbols *, ** and ***, respectively.

**Results**

In Table 1 relevant terms from climate science (IPCC, 2013) are matched to terms for control systems (Astrom and Murray, 2008). In this study, the control system *process variable* is specified as global surface temperature. Next, it is accepted (IPCC, 2013) that the main driver of temperature is atmospheric $CO_2$. Hence, if levels of the driver of temperature change markedly, this can, if nothing intervenes, cause perturbation of temperature. So in what follows we will use the level of $CO_2$ time series to represent the control system term *disturbance*.

Table 1.

| Climate science term (IPCC, 2013) | Control systems term (Astrom and Murray, 2008) | | As operationalised | Variable observed | Variable transformed | Comment on transformation |
|---|---|---|---|---|---|---|
| Temperature | Process variable | | | Observed (global surface) temperature (Temp) | Z_Temp | |
| | Setpoint | | | Initial temperature of series | Z_Temp_setpoint | Setpoint is set up to equal zero |
| Driver | Disturbance | | | Level of atmospheric $CO_2$ ($CO_2$) | Z_ $CO_2$ | |
| | | Error | Error = setpoint - disturbance | | Z_Temp_setpoint minus Z_ $CO_2$ | As setpoint is defined as zero, Z_Temp_setpoint minus Z_ $CO_2$ reduces simply to minus Z_ $CO_2$ |
| | Controller output | Reverse error | | | Z_ $CO_2$ | |
| | Proportional Controller output | Controller output | | | Z_ $CO_2$ | |

| | Integral Controller output | Integral Controller output | Cumulative sum of Controller output | | I_Z_ $CO_2$ | |
| | Derivative Controller output | Derivative Controller output | First difference of Controller output | | d_Z_ $CO_2$ | |

After Astrom and Murray (2008), and as shown in Table 1, the *error* term determined by the control system equals setpoint minus disturbance.

As stated above, feedback control systems (Astrom and Murray, 2008) derive an error term or terms which are fed back as reactions to disturbances to the system with the aim of keeping the system at its previously determined setpoint.

We therefore express this relationship in terms of an equation as:

Eqn 1. Setpoint = Disturbance - P_error - I_error - D_error (each term with its respective coefficient)

With regard to controller terminology, this is equivalent to:

Eqn 2. Setpoint = Disturbance + P_controller_output + I_controller_output + D_controller_output

For simplicity in what follows we will use the expression in Eqn 1. For a temperature control system, then, Eqn 1 becomes:

Eqn 3. Setpoint temperature = Disturbance - P_error - I_error - D_error

If the temperature control system does not reach its setpoint temperature this can be seen to be expressed as:

Eqn 4. Observed temperature = Disturbance - P_error - I_error - D_error

In more detail regarding equations 3 and 4, in this study we might seek to correlate the sought outcome (dependent) variable, setpoint, with the disturbance variable and the three P, I and D error variables. However as the P_error time series by definition is the precise opposite mathematically of the Disturbance time series this would result in setpoint equalling Disturbance plus P_error leaving no role for I_error and D_error.

However it can be argued that the complexity of having I and D controller outputs in 'real world' control systems such as industrial control systems is entirely because real world control is not as simple as setpoint equalling Disturbance minus P_controller_output.

What we use therefore as the outcome variable is not the setpoint temperature, but *the real world result of the attempt of the system to seek and attain the setpoint temperature* – this is the observed global surface temperature. Further, if a PID system is operating, signatures of the effect of each of its P, I and D controller output variables on the disturbance variable – in seeking to achieve the setpoint outcome – should be discernable in the observed temperature series.

Turning to setpoint, because in this instance the setpoint equals zero due to Z-scoring, the error term becomes simply zero minus disturbance, which equals minus disturbance.

The I and D error terms are derived from the P error term as the cumulative sum of the disturbance series, and the first difference of the disturbance series respectively.

The blue curve in Figure 1 shows the disturbance to temperature, the level of atmospheric $CO_2$. We do not know the actual setpoint of the putative control system. However, the level of $CO_2$ (disturbance) curve (blue curve) is steadily rising (the disturbance is getting worse). Hence, any point towards the start of the data series must be closer to the level of the temperature setpoint. We therefore choose a value for the setpoint (i) earlier in the time period; and (ii) as the average of a period over which the temperature was roughly level – this is the period from the start of data in 1959 to 1976. All curves in the figure are therefore Z-scored using a 1959 to 1976 base period. This enables us to see any points of departure thereafter. If there is no change in relationships between the curves, their trajectories will remain common after 1976, just as they were before. The setpoint is shown as the yellow curve. The observed temperature is also plotted (purple line).

Figure 1. Monthly data, Z-score base period 1958-1976 (period to left of vertical black line). Putative control system model for global surface temperature, selected elements: Disturbance (level of atmospheric $CO_2$) (black curve); control system setpoint for temperature (purple curve); observed temperature (red curve).

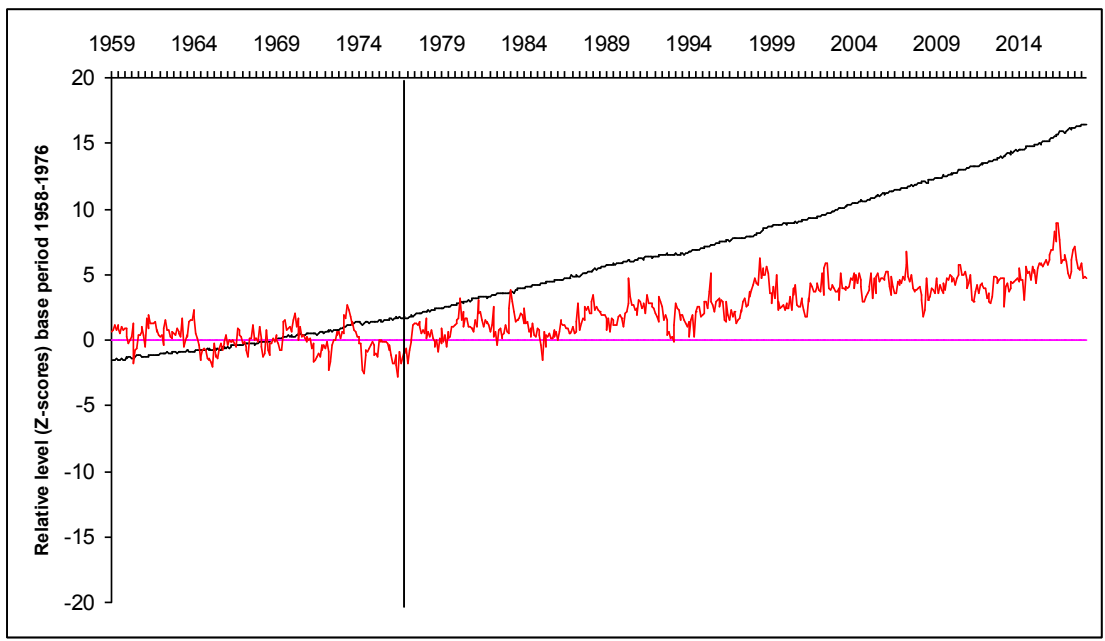

It can be seen that the disturbance curve has risen above the setpoint value selected. The observed temperature has also risen above the setpoint since 1976, but has stayed nearer to the setpoint than to that which might be predicted from the disturbance.

Figure 2 adds the P, I and D error terms to the above curves.

Figure 2. . Monthly data, Z-score base period 1958-1976 (period to left of vertical black line). Putative control system model for global surface temperature, full set of elements: Disturbance (level of atmospheric $CO_2$) (black curve); control system setpoint for temperature (purple curve); observed temperature (red curve); level of control system error (P_error) ((yellow curve) integral (cumulative sum) of control system error (I_error) (orange curve); derivative (first difference) of control system curve (D_error)(brown curve)

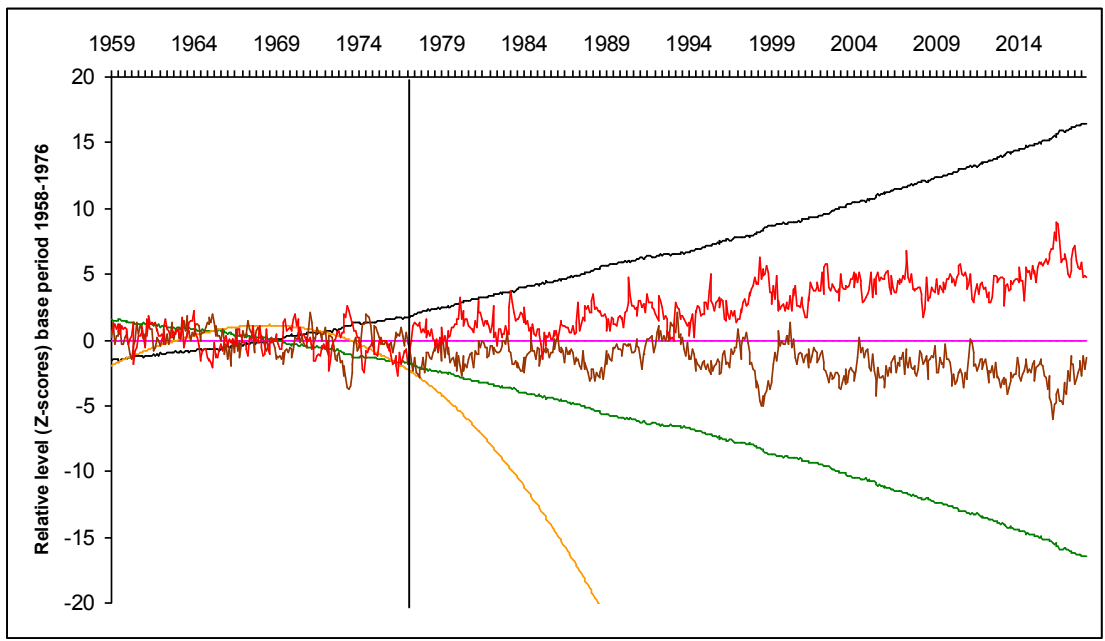

As expected, it can be seen that each error term, if applied, would tend to lower the temperature below that implied from the trend of the disturbance.

We now turn to the time series analysis of the relationship of global surface temperature to this putative control system and its proportional, integral and derivative feedback components.

First we assess the order of integration of the series used.

For each series ,Table 2 displays the results of tests for stationarity, allowing for both drift and trend.

Table 2. Results of ADF tests for stationarity, allowing for both drift and trend.

|  | Disturbance series | P_error series | I_error series | D_error series | Temp series |
|---|---|---|---|---|---|
| Order of polynomial trend determined for use in constructing the test | Same as P-error series | 4$^{th}$-order | 3$^{rd}$-order | Linear | Linear |
| Test indicated for use by order of polynomial trend of series | | Schmidt-Phillips test (τ test) | Schmidt-Phillips test (τ test) | Augmented Dickey-Fuller test | Augmented Dickey-Fuller test |
| Value of test-statistic | | -7.6313 | -9.1785 | -5.716 | -0.17473 |
| Critical value of test statistic for a significance level of 0.01 | | -4.85 | -4.5 | -3.9713 | -7.03966 |
| Conclusion: series is – | | Stationary | Stationary | Stationary | Stationary |

Table 2 shows that all series are stationary.

We now commence regression analysis. As all series are stationary, OLS can be used, with ARIMA dynamic regression analysis used for any autocorrelation detected.

Both ridge regression and ARDL dynamic regression results are provided.

Although the ridge regression process that we use (Schreiber, 2017) automatically Z-scores the series used as an input to its process, the output is then also automatically rescaled to match the original data. To enable the comparison of the strength of each coefficient in the ridge regression (see Methods), we need input series Z-scored using the whole time period 1958 to 2018 as the base period, so this is employed. Next, P, I and D error terms are prepared from these $CO_2$ series, Z-scored from 1958 to 2018. Third, once prepared, the resulting P, I and D error term series are re-Z-scored using the same period 1958 to 2018. These series are then used for the ridge regression.

Tables 3 to 5 now provide ridge regression and ARDL results for the relationship of the putative PID control system with temperature.

Table 3. Ridge regression results: putative PID control system

|  | coefficient | std. error | z | p-value |
|---|---|---|---|---|
| Disturbance_Z5818 | 0.2630 | 0.0202 | 13.03 | 7.87E-39 |
| P_Error_reZ5818 | -0.2630 | 0.0202 | -13.03 | 7.87E-39 |
| I_Error_reZ5818 | -0.2174 | 0.0369 | -5.897 | 3.69E-09 |
| L1m_13mma_D_Error_reZ5818 | -0.2377 | 0.0218 | -10.9 | 1.15E-27 |
| const | 0 | NA | NA | NA |

Penalty = 0.57  
Resid SD = 0.44516

All three P, I and D terms display coefficients of the same order of magnitude in the ridge regression.

To gain information on statistical significance, we now turn to enabling standard OLS by means of the variable subset selection method. The collinear variable deleted to produce the subset is Disturbance. Dynamic OLS regression is then conducted using the ARDL package.

Table 4. Eviews ARDL estimation output for period 1958 to 2018 for temperature as a function of putative control system P_error, I_error and D_error: short-run model dynamic relationship

| Model dependent variable | Number of models evaluated | Selected model | No autocorrelation out to 36 lags | Adjusted R-squared | F-statistic | p-value |
|---|---|---|---|---|---|---|
| H46_Z5876 | 2 | ARDL (2,0,0.0) | Nil | 0.896911 | 1234.712 | < 10-100 |

Table 5. Eviews ARDL estimation output for period 1958 to 2018 for temperature as a function of putative control system P_error, I_error and D_error: short-run model independent variables

| Model independent variables | Coefficient | Std. Error | t-Statistic | Prob.* |
|---|---|---|---|---|
| H46_Z5818(-1) | 0.523 | 0.0368 | 14.222 | 1.57E-40 |
| H46_Z5818(-2) | 0.226 | 0.0367 | 6.162 | 1.21E-09 |
| P_ERROR_REZ5818 | -0.139 | 0.0334 | -4.158 | 3.60E-05 |
| I_ERROR_REZ5818 | -0.046 | 0.0277 | -1.675 | 0.0944 |
| L1M_13MMA_D_ERROR_REZ581 | -0.062 | 0.0173 | -3.596 | 0.0003 |
| C | 0.002 | 0.0121 | 0.174 | 0.8621 |

Although the significance of the I-error term is only at the 0.1 level, all three P, I and D terms are significant in the model.

It is still possible, however, that other control system models using only P, I, or D, or PI or PD terms might display even better fits. The same assessment process as above was carried out for these models and the results are shown in Table 6 and Figure 3.

Table 6. Fit with observed temperature of global surface temperature predicted from dynamic regression models involving various combinations of control system error terms.

| Error terms in model | Selected model | No autocorrelation out to: | Adjusted R-squared | Akaike information criterion |
|---|---|---|---|---|
| I | (2,0) | 36 lags | 0.8921 | 0.6188 |
| D | (4,0) | 37 lags | 0.8930 | 0.6152 |
| P | (2,0) | 36 lags | 0.8951 | 0.5900 |
| ID | (2,0,0) | 24 lags | 0.8945 | 0.5972 |
| PI | (2,0,0) | 36 lags | 0.8952 | 0.5912 |
| PD | (2,0,0) | 36 lags | 0.8966 | 0.5769 |
| PID | (2,0,0,0) | 36 lags | 0.8969 | 0.5758 |

Figure 3. Fit with observed temperature of global surface temperature predicted from dynamic regression models involving various combinations of control system error terms. The PID model shows the best fit.

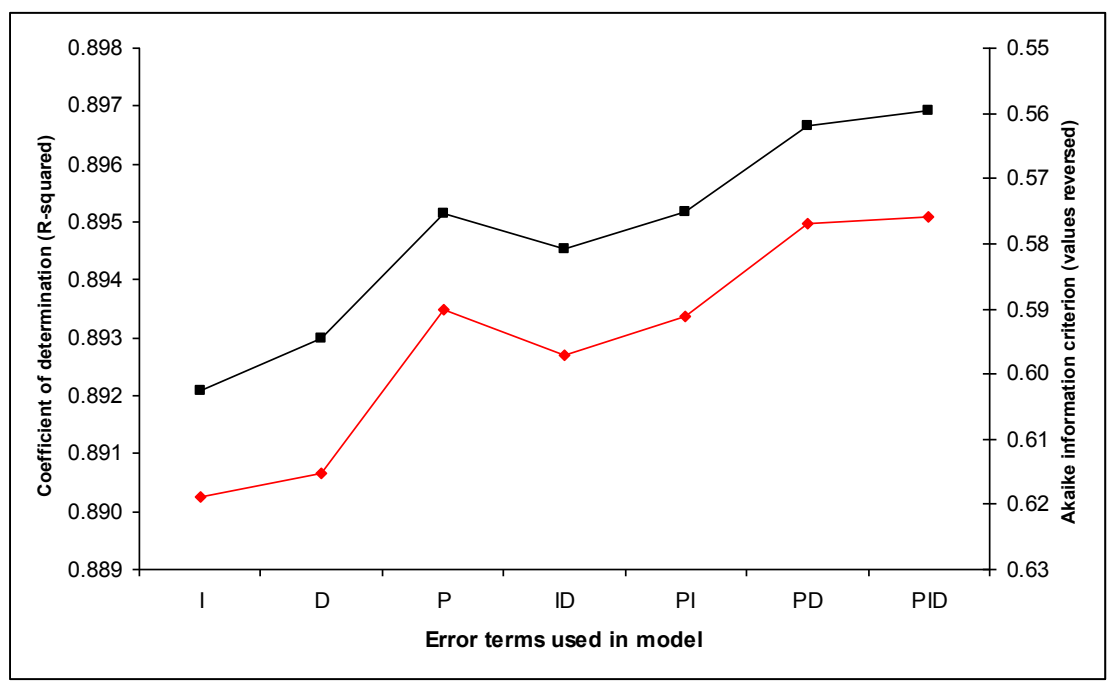

Table 6 and Figure 3 show that, against the other models, a model involving all three P, I, and D control outputs displays the highest R-squared value, and the best (lowest) Akaike information criterion value. This is evidence that the control system is a full PID control system, rather than one using fewer control outputs.

To see the effect of the control system over time, and against the alternative falsifying hypothesis (Kirchner, 1989) of no control system existing, we now return to series Z-scored over the earlier part of the model – 1958 to 1976 – as for Figures 1 and 2.

These results are depicted in Figure 4.

Figure 4. Monthly data, Z-score base period 1958-1976 (period to left of vertical black line). Predicted temperature from dynamic regression model of a control system for global surface temperature using P, I and D error terms (red curve); Disturbance to temperature (level of atmospheric $CO_2$) (black curve); observed temperature (green curve);

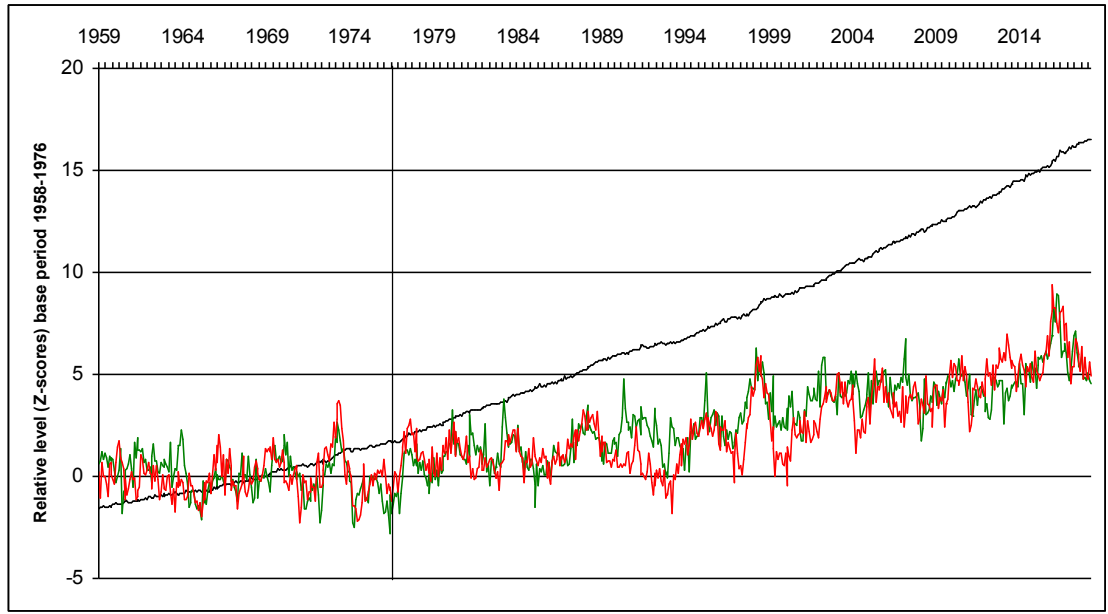

The figure shows that there is a close match between the temperature expected from the control system model (yellow curve) and the observed temperature (purple curve). This is the case especially when compared to the alternative falsifying hypothesis of 'no control system' which would imply that the disturbance driver would be closely followed (blue curve).

Figure 4, then, provides support for the hypothesis of a control system, and no support for the falsifying prediction of the absence of a control system.

**Discussion**

In the foregoing analysis, we made a case that the outcome variable to use is not the setpoint temperature, but the real world result of the attempt of the system to meet the setpoint temperature – the observed global surface temperature. We argued that if a PID control system is operating to achieve a setpoint outcome, signatures of the effect of each of its P, I and D controller output variables on the disturbance variable should be seen in the outcome actually occurring, the observed temperature series.

We consider our results using observed data at global scale show those precise signatures, revealed as highly statistically significant relationships of PID form. We believe this result to be comprehensive and suggestive (and perhaps even diagnostic) of the existence of a Gaia-type control system presently operating at global scale on global surface temperature.

With regard to the robustness of the results, we first turn to criteria for the specification of hypotheses and their testing. Here we believe the results of our analysis are in a form meeting the criteria of a major critic of Gaia, Kirchner (1989).

As mentioned in the Introduction, Kirchner (1989) set out three criteria:

> In order to be testable, a hypothesis must be clear, and its terms must be unambiguous;
> It must be intelligible in terms of observable phenomena;
> It must generate predictions of two kinds: confirmatory predictions (phenomena that should be observed if the hypothesis is true and that would not be predicted by the existing body of accepted theory) and falsifying predictions (phenomena that should be observed if the hypothesis is false)

Our analysis has used observed phenomena throughout. The prediction of a PID control system was specified, and evidence for a PID control system was then sought and found.

We consider that finding no evidence for a PID control system in our study would constitute a form of falsifying prediction

Next, in the Introduction, we noted that on conceptualising Gaia, Kirchner (2002, 2003) referred to three major themes of the Gaia Hypothesis and provided critiques of them: Gaia and Homeostasis, Gaia and Environmental Enhancement, and Gaia and Natural Selection. The first two themes we considered germane to this paper.

With respect to Gaia and Homeostasis, Kirchner asked whether biological feedbacks stabilise the global environment: that is, whether the 'Homeostatic Gaia' hypothesis is correct

Kirchner (2003) argued that: "…in the context of the past 30 years …Gaia's proponents have vigorously advanced the view that life stabilizes Earth's climate, while largely ignoring the substantial body of evidence that biological feedbacks also can (and do) destabilize Earth's climate."

We considered that the appropriate approach was to look at a *global level* performance metric and see if its trend is towards stabilisation or destabilisation.

Our results show that, as the PID system keeps the global surface temperature lower than expected from the disturbance trend and nearer to the setpoint, the trend of temperature is towards stabilisation. Hence, for temperature, we consider our results show that Kirchner's (2003) question of the net of stabilising and destabilising feedbacks is resolved in favour of stabilisation.

We now turn to Kirchner's Gaia and Environmental Enhancement theme, which he described as "…the claim that the biota alter the physical and chemical environment to their own benefit."

To the extent that the current level of the $CO_2$ driver is higher than it has been over the past 800,000 years, and temperature tracking that driver is expected to produce threats to the biosphere (IPCC 2013), any temperature lower than that forecast by the trend in the $CO_2$ driver and keeping closer to the prior level can be considered beneficial. Hence, for temperature, we consider our results show that Kirchner's (2003) question of whether "the biota alter the physical and chemical environment to their own benefit" is answered in the affirmative to the extent that our evidence suggests that a global control system is altering global temperature to the benefit of the biota.

One outstanding question is where this planet-scale control system might reside. The control system is acting on temperature and making things cooler than they would otherwise be. There is evidence that the oceans cool (IPCC, 2013) and that evapotranspiration (which is dominated by plants) adds to this (Leggett and Ball, 2018).

While beyond the scope of the present paper to further explore the location of the planet-scale control system, suffice it to say that a highly dynamic PID-type process at monthly level is hard to expect to originate from the inanimate elements of the ocean (and land). By contrast, the processing of information required for a control system is quite within the capability of plants. Indeed Spitzer and Sejnowski (1997) argue that rather than occurring rarely, differentiation and other computational processes are present and potentially ubiquitous in living systems, including even at the single-celled level where a variety of biological processes – concatenations of chemical amplifiers and switches – can perform computations such as exponentiation, differentiation, and integration.

**References**

Ahmad, N., & Du, L. Effects of energy production and $CO_2$ emissions on economic growth in Iran: ARDL approach. *Energy* **123**, 521–537 (2017).

Allen, M.P. *Understanding regression analysis* (Plenum, New York, 1997).

Amemiya, T. *Advanced econometrics* (Harvard University Press, Cambridge, 1985).

Araki, M. PID control. *Control systems, robotics and automation*, **2** 1-23 (2002).

Åström, K. J. & Murray, R. M. *Feedback Systems: An Introduction for Scientists and Engineers* (Princeton University Press, Princeton, 2008).

Birkes, D., & Dodge, Y. *Alternative Methods of Regression* (Wiley, New York, 1993).

Bogen, J. Theory and observation in science. in *The Stanford Encyclopedia of Philosophy*. (2017). Retrieved from http://plato.stanford.edu/archives/spr2018/entries/science-theoryobservation

Dyke, J.G. & Weaver, I.S. The emergence of environmental homeostasis in complex ecosystems. *PLoS Comput. Biol.* **9**, e1003050 (2013).

Elliott, G., Rothenberg, T. J. & Stock J. H. Efficient tests for an autoregressive unit root. *Econometrica*, **64**, 813–836 (1996).

Enting, I.G. A modelling spectrum for carbon cycle studies. *Math Comput Simul* **29**, 75–85 (1987).

Farrar, D. E. & Glauber, R. R. Multicollinearity in regression analysis: the problem revisited. .*Review of Economics and Statistics* **49**, 92-107 (1967).

Granger, C. W. J. & Newbold, P. Spurious regressions in econometrics. *J. Econom.* **2**, 111–120 (1974).

Greene, W. H. *Econometric Analysis* (Prentice Hall, Boston, 2012).

Goodwin, G. C., Graebe, S. F. & Salgado, M. E. *Control System Design* (Prentice Hall, Englewood Cliffs , 2000).

Greene, W. H. *Econometric Analysis* (Prentice Hall, Boston, 2012).

Gretl 2018a *GNU Regression, Econometrics and Time-series Library* available at: https://sourceforge.net/projects/gretl/ (last accessed 20 August 2018).
Harvey, D. I., Leybourne, S. J. & Taylor, A. M. R. Testing for unit roots and the impact of quadratic trends with an application to relative primary commodity prices. *Discussion Paper No. 08/04, Granger Centre for Time Series Econometrics, University of Nottingham* (2008).

Hastie T., Tibshirani R. &Friedman J. *The elements of statistical learning: data mining, inference, and prediction (*Springer, New York, 2009).

Hoerl, A. E. Application of ridge analysis to regression problems. *Chemical Engineering Progress* **58**, 4-59 (1962).

Hoerl, A. E. & Kennard, R. W. Ridge regression: applications to nonorthogonal problems. *Technometrics* **12**, 69–82 (1970).

IHS EViews: EViews 9.5, IHS Global Inc., Irvine, California, 2017.
available at: http://www.eviews.com/download/download.shtml
(last accessed 20 August 2018).

IPCC *Climate change 2013: the physical science basis. Contribution of working group I to the fifth assessment report of the intergovernmental panel on climate change* (eds Stocker T.F., *et al.* ) (Cambridge University Press, Cambridge 2013).

Janjua, P.Z., Samad, G. & Khan, N. Climate change and wheat production in Pakistan: an autoregressive distributed lag approach. *NJAS-Wagening J. Life Sci.* **68**, 13–19 (2014).

Karplus, W.J. The spectrum of mathematical modelling and systems simulation. *Math. Comput. Simul.* **19**, 3–10 (1977).

Karplus, W.J. *The Heavens are Falling: The Scientific Prediction of Catastrophes in our Time* (Plenum, New York, 1992).

Keeling, R.F., Piper, S.C., Bollenbacher, A.F. & Walker, S.J. (2009) Carbon Dioxide Research Group, Scripps Institution of Oceanography (SIO), University of California, La Jolla, California USA 92093- 0444, available at: http://cdiac.ornl.gov/ftp/trends/CO2/maunaloa.CO2 (last accessed 10 August 2018).

Kirchner, J. W. The Gaia hypothesis: can it be tested? *Rev. Geophys* **27**, 223–235 (1989).

Kirchner, J. W. The Gaia hypotheses: are they testable? Are they useful? in *Scientists on Gaia* (ed Schneider, S. H. & Boston, P. J.) 38-46 (MIT Press, Cambridge, 1991).

Kirchner, J. W. The Gaia hypothesis: fact, theory, and wishful thinking. *Clim. Change* **52**, 391–408 (2002).

Kirchner, J. W. The Gaia hypothesis: Conjectures and refutations. *Clim. Change* **58**, 21–45 (2003).

Kleidon, A. Testing the effect of life on Earth's functioning: How Gaian is the Earth system? *Clim. Change* **52**, 383–389 (2002).

Leggett, L.M.W. & Ball, D.A. Granger causality from changes in level of atmospheric $CO_2$ to global surface temperature and the El Niño–Southern Oscillation, and a candidate mechanism in global photosynthesis. *Atmos. Chem. Phys.* **15**, 11571–11592 (2015).

Leggett, L.M.W. & Ball, D.A. Evidence that global evapotranspiration makes a substantial contribution to the global atmospheric temperature slowdown. *Theor. Appl. Climatol.* (2018).

Lenton, T. M. Testing Gaia: the effect of life on Earth's habitability and regulation. *Clim. Change* **52**, 409–422 (2015).

Lenton T.M. & Wilkinson, D.M. Developing the Gaia theory. *Clim. Change* **58**, 1-12 (2003).

Lovelock, J.E. & Margulis, L. Atmospheric homeostasis by and for the biosphere: the Gaia hypothesis. *Tellus Series A* **26** , 2–10 (1974).

Lovelock, J. The living Earth. *Nature* **426**, 769-770 (2003).

Moody, D. Seven misconceptions regarding the Gaia hypothesis. *Climatic Change* **113**, 277–284 (2012).

Morice C.P., Kennedy J.J., Rayner N.A. & Jones P.D. Quantifying uncertainties in global and regional temperature change using an ensemble of observational estimates: The HadCRUT4 data set. *J. Geophys. Res. Atmos.* **117** (2012) HadCRUT4 data used available at: http://www.metoffice.gov.uk/hadobs/hadcrut4/data/current/time_series/HadCRUT.4.4.0.0.monthly_ns_avg.txt, last access: 12 August 2018.

Nisbet, E. G., Fowler, C. M. R. and Nisbet, R. E. R. The regulation of the air: a hypothesis. *Solid Earth* **3**, 87–96 (2012).

Pesaran, M.H., Shin, Y. & Smith, R.J. Bounds testing approaches to the analysis of level relationships. *J. Appl. Econ.* **16**, 289–326 (2001).

Pfaff, B., Zivot, E. & Stigler, M.. Package urca. Unit root and cointegration tests for time series data. https://cran.r-project.org/web/packages/urca/urca.pdf2016. Accessed September 2018.

Phillips, P.C.B. Understanding spurious regressions. *Journal of Econometrics* **33**, 311-340 (1986).

R Development Core Team. R: a language and environment for statistical computing. R Foundation for Statistical Computing, Vienna, Austria. (2009). Downloaded from https://mirror.its.sfu.ca/mirror/CRAN/ Accessed September 2018

Schreiber, S. Notes on Gretl Ridge regression function (2017), accessed July 2018.

Sackett, D.L., Rosenberg W.M., Gray J.A., Haynes R.B. & Richardson WS. Evidence based medicine: what it is and what it isn't. *BMJ* **312**, 71-72 (1996).
Schmidt, P. & Phillips, P. C. B. LM test for a unit root in the presence of deterministic trends. *Oxford Bulletin of Economics and Statistics*, **54**, 257–287 (1992).

Spitzer, N. C., & Sejnowski, T. J. Biological information processing: bits of progress, *Science* **277**, 1060–1061 (1997).

Tyrrell, T. *On Gaia: A critical investigation of the relationship between life and Earth* (Princeton University Press, Princeton, 2013).

VanDoren, V.J. Basics of proportional-integral-derivative control. *Control Engineering* **45,** 135-145 (1998).

Wilkinson, D.M. On Gaia: a critical investigation of the relationship between life and Earth. *Int. J. Environmental Studies* **72,** 724-730 (2015).

| | |
|---|---|
| Filename: | Leggett and Ball.doc |
| Directory: | C:\LEXAR\Global warming |
| Template: | C:\Users\User\AppData\Roaming\Microsoft\Templates\Normal.dot |
| Title: | Many processes in the Earth's surface essential for the conditions of life depend on the interaction of living forms, especially microorganisms, with inorganic elements |
| Subject: | |
| Author: | TLM-User |
| Keywords: | |
| Comments: | |
| Creation Date: | 28/09/2018 2:48:00 AM |
| Change Number: | 2 |
| Last Saved On: | 28/09/2018 2:48:00 AM |
| Last Saved By: | TLM-User |
| Total Editing Time: | 1 Minute |
| Last Printed On: | 2/10/2018 6:39:00 AM |

As of Last Complete Printing
    Number of Pages: 23
    Number of Words:    8,271 (approx.)
    Number of Characters:    47,145 (approx.)